\documentclass[pra,preprint]{revtex4}
\usepackage{graphicx}
\usepackage [english]{babel}
\usepackage{amssymb}
\usepackage{amsmath}
\usepackage{amsthm}

\begin{document}

\title{\Large{\bf{The role of the slope of `realistic' potential barriers
\\
in preventing  relativistic  tunnelling in the Klein zone}}}

\author{Paolo Christillin}
\email{christ@df.unipi.it}
\author{Emilio d'Emilio}
\email[Corresponding author: ]{demilio@df.unipi.it}
\affiliation{Dipartimento di Fisica, Universit\`a di Pisa}
\affiliation{
Istituto Nazionale di Fisica Nucleare, Sezione di Pisa, Pisa,
Italy}

\date{\today}



\begin{abstract}

The transmission of fermions of mass $m$ and energy $E$ through an 
electrostatic potential barrier of rectangular shape (i.e. supporting an {\it infinite} electric field), of
height
$U> E+ m\,c^2$ - due to the many-body nature of the Dirac equation  evidentiated by the Klein paradox - has
been widely studied. We exploit here the analytical solution, given by Sauter for the linearly rising
potential step, to show that the tunnelling rate through a more realistic trapezoidal barrier
is exponentially depressed, as soon as the length of the regions supporting a
{\it finite} electric field exceeds the Compton wavelenght of the particle -
the latter circumstance being hardly escapable in most realistic cases.

\vglue 1 cm

\hfill{PACS numbers 03.65.Pm, 03.65.Xp}

\end{abstract}

\maketitle


\section{Introduction and main result}    \label{intro}
\begin{figure}
              \includegraphics{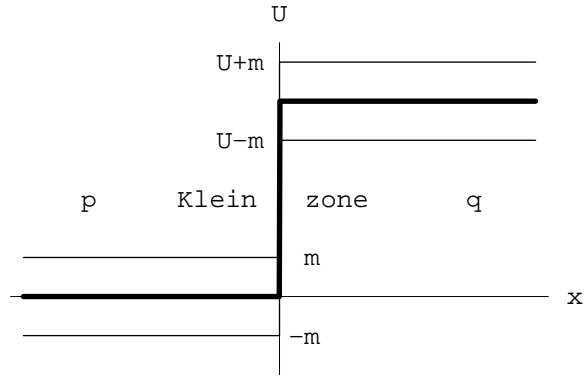}
              \vspace{.1 cm}
\caption{Klein step} \label{fig: Fig1}
\end{figure}

We will consider a one-dimensional flux of free monoenergetic electrons of 
 energy $E$ and momentum 
\begin{eqnarray}
p=\sqrt{E^2-m^2}
\label{p}
\end{eqnarray}
(throughout the paper natural units $\hbar=c=1 $ will be used)
hitting a repulsive rectangular potential step of height $U$ greater than their kinetic energy $E-m$
(Fig. \ref{fig: Fig1}) - to which they are {\it minimally} coupled. In this case  total
reflection is unavoidable: in the region $x>0$ there is indeed just one bounded solution of
the Schr\"odinger 
equation and the flux is therefore zero.
As a consequence, in the region
$x<0$ the reflected flux must equal the incident one.

The use of the Schr\"odinger equation is legitimate as long as both $U,\,E-m\ll m$ and,
in principle, the result must not necessarily hold for higher values of $U$.
Indeed for 
\begin{equation}
U-m> E > m \qquad{\rm (Klein \ zone)}
\label{KZ}
\end{equation}
(the only zone we will be interested in throught the paper)
a relativistic equation is more suited for the description of the situation and, for
both the Klein-Gordon and the Dirac equation (to which our discussion will be limited), it
happens that in the region $x>0$ the plane-wave (free propagation) solutions with opposite
momenta
$\pm q$,
\begin{equation}
q= \sqrt{ (U-E)^2-m^2 }
\label{q}
\end{equation}
are two. Therefore the possibility of a non trivial transmitted flux of the same order of
magnitude as the incident one is re-opened. The first to point out such
a seemingly paradoxical result was Klein \cite{KL} in the case of the then recently
proposed Dirac equation.
\begin{figure}
              \includegraphics{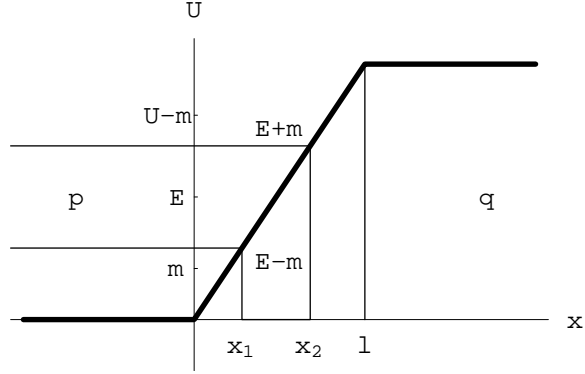}
              \vspace{.1 cm}
\caption{Sauter step} \label{fig: Fig2}
\end{figure}
The result has been commented upon and used by several authors over the years: see e.g.
ref. \cite{CalDom} for a historical review and some references.  For us it is important to mention that
Klein's result was questioned by Sauter \cite{SA} who, following  a suggestion by Bohr, showed - at least
for the Dirac equation - that, when the sharp 
edge of the step is substituted by a
more realistic one of width $\ell$ (FIG. \ref{fig: Fig2}), the transmission coefficient turns out to be 
\begin{equation}
T^{\rm Sauter \ step}_{\rm D}\simeq e^{-\pi\, m^2\, \ell/U}\ .
\label{TSau}
\end{equation}  
The asymptotic form exhibited in (\ref{TSau}) holds when both the particles are fast $E-m\simeq m$ and the
slope of the step, i.e. the electric field, satisfies
\begin{equation}
e\,{\cal E} = U/\ell\ll 2m/\lambda_{\rm
Compton}= 2 m^2\,.
\label{Erestr}
\end{equation} 
A discussion, in the above framework, of whether the Klein paradox is a `real' one  and - at least for the
Dirac equation - Sauter's is the right way out, cannot help spelling out what particles do the asymptotic
states describe. Indeed, while in the region
$x< 0$ the dispersion relation
$E=+\sqrt{p^2+m^2}$ is unambiguous, in the region $x> \ell$, owing to (\ref{KZ}) and (\ref{q}),
either
\begin{equation}
E = U -\sqrt{q^2+m^2}
\label{electrons}
\end{equation}  
or
\begin{equation}
-E = -U + \sqrt{q^2+m^2}\ .
\label{positrons}
\end{equation} 
Namely either particles have negative kinetic energies, or antiparticles (propagating backward
in time) have to come into play: the first choice being untenable, the many-body nature of
both relativistic equations can no longer be ignored.

Postponing the discussion of this point until Section II, we prefer instead to
examine the Klein paradox in the context of  a different `Gedanken' experiment
where it is possible, for a while, to ``sweep the dust under the rug": can an appreciable
fraction of monoenergetic electrons pass through a potential {\it barrier} of height  $U\agt 2m\,$?

\begin{figure}
              \includegraphics{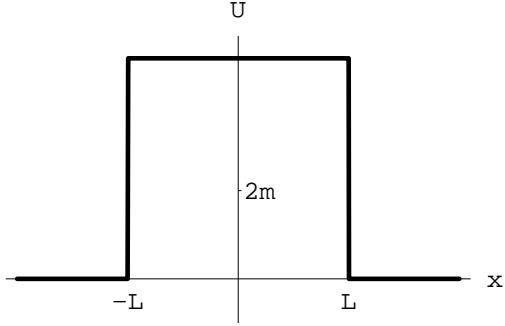}
\caption{Klein barrier} \label{fig: Fig3}
\end{figure}

\begin{figure}
              \includegraphics{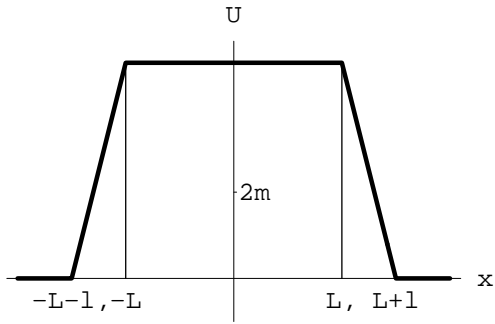}
              \vspace{.1 cm}
\caption{Sauter barrier} \label{fig: Fig4}
\end{figure}

The situation is
summarized in FIG.s \ref{fig: Fig3} and \ref{fig: Fig4}. The former is the well studied rectangular
barrier of width $2L$ whose transmission coefficient is known both in the Klein-Gordon and Dirac
case:
\begin{equation}
T_\pm^{\rm rect \ bar}= \frac{1}{1+ f_\pm(E,U)\,\sin^2 (2 q \,L)}
\label{rectBarr}
\end{equation} 
\begin{equation}
f_\pm(E,U) = 
\begin{cases}
\displaystyle{\frac{U^2(U/2-E)^2}{p^2 \, q^2}} &  \qquad   +=  \text{KG} 
                \\  \\
\displaystyle{\frac{m^2\,U^2}{p^2 \, q^2}} & \qquad  -=  \text{D} 
\end{cases}
\label{KGD}
\end{equation} 
$p, \ q$ being given by (\ref{p}), (\ref{q}) respectively. This is the schematization of `infinite
electric field': the width of the edges of the barrier is neglected with an ensuing electric field 
 much higher then its relevant scale $2m^2/e$ given in (\ref{Erestr}). This
limitation is instead removed in the case of the trapezium shaped potential of Fig. \ref{fig: Fig4}, the
extension of Sauter's cure to the barrier. It is our choice to preserve space inversion as
a symmetry (it will be evident that our conclusion does not critically depend
on this assumption) and we will nickname this potential as Sauter barrier. 

The advantage of barriers with respect to
steps is that in both the regions where $U=0$, one may  choose asymptotic states
describing particles and ignore  `what is going on under
the barrier' (we mean $-L<x<L$), i.e. whether either particles or antiparticles and/or
couples are freely propagating.

This aspect has been examined in \cite{KSG} for the step, in \cite{dLR} for the barrier, by adopting a
space-time description of the scattering process instead of the stationary state picture we are
sticking to here.

\begin{figure}
              \includegraphics{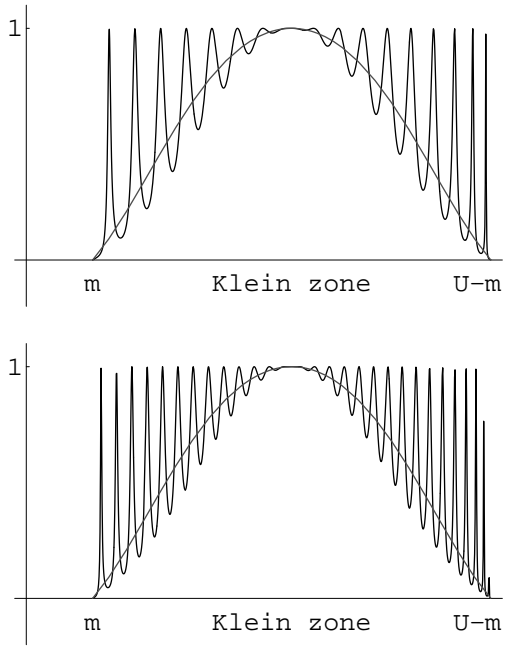}
              \vspace{.1 cm}
\caption{$T^{\rm rect \ barr}_{\rm KG}$ as a function of $E$ in the Klein zone:
$U=8m$, $L=4/m$ (up) and $L=6.5/m$ (down)} 
\label{fig: Fig5}
\end{figure}

\begin{figure}
              \includegraphics{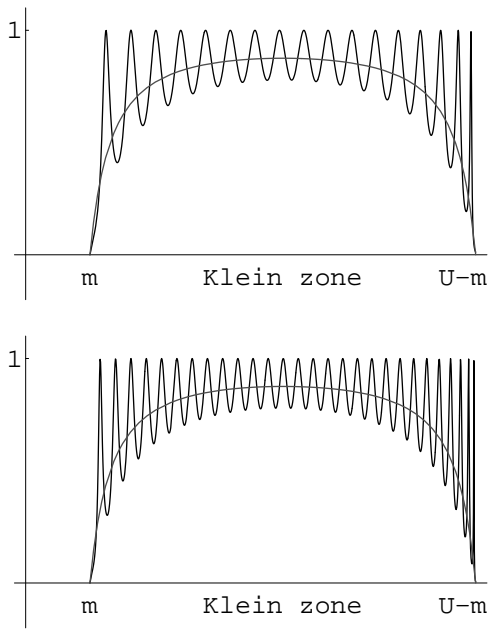}
              \vspace{.1 cm}
\caption{$T^{\rm rect \, barr}_{\rm D}$ as a function of $E$ in the Klein zone:
$U=8m$, $L=4/m$ (up) and  $L=6.5/m$ (down)} 
\label{fig: Fig6}
\end{figure}

Also the authors of \cite{dLR} claim
a `barrier paradox'.
How does the paradox show up in the case of the rectangular barrier?
Firstly it will be noted in (\ref{rectBarr}) that for the values of $E$ such that $2q\,L=n\, \pi$, $n$
integer, there is total transmission (the so-called transmission resonances), both in the
Klein-Gordon and the Dirac case, no matter how high the values of $U$ and $L$ may be. This is  in
blatant contrast with the Schr\"odinger case mentioned in the beginning.
A second crucial
difference is in the role played by the barrier width parameter $L$. While its increase
induces the well known exponential decrease of 
$T^{\rm rect \ bar}_{\rm Schroedinger}$, in both the relativistic cases - for fixed
$U$ - the higher the value of
$L$, the faster the oscillations of $T_\pm^{\rm rect \ bar}$ as a function of $E$. Fig.s \ref{fig: Fig5}
and \ref{fig: Fig6} illustrate this statement.

To make our point we have now to take into account how a `realistic'
experiment would be carried out and to give the numerical values
of the scales involved. 


Taking
$L\gg 1/m= \lambda_{\rm Compton}= 0.024$ \AA  \ for electrons (and smaller for more massive bosons) is simply
mandatory.  In addition, in a `real experiment', where almost monoenergetic
particles - with average energy $\overline E$ and an energy uncertainty  $\Delta E\ll
\overline E$ - are gunned against the barrier, the in-state is obtained as some
superposition of stationary states with $\overline E -\frac{1}{2}\Delta E \alt E \alt \overline E
+\frac{1}{2}
\Delta E$. Thus, even for reasonably monoenergetic electrons, a huge number of transmission
resonances is involved in the diffusion process.
As an example, for
$m=0.51$ MeV,
$U\ge 1.2$ MeV,
$L \ge  10$ \AA \ and  
$\overline E\simeq m+ \frac{1}{2}\Delta E = 0.520$ MeV  (i.e. at the very beginning of the Klein
zone with a relative monoenergeticity $\Delta E/\overline E \le 4\%$) no less than 1450 spikes in
the analogues of Fig.s \ref{fig: Fig5} and \ref{fig: Fig6} are involved. 

Such fast oscillations have to be taken into account \cite{CalDom}. This is done by
replacing the
$\sin^2(\dots)$ with  $\frac{1}{2}$ in (\ref{rectBarr}) and this, in turn, leads to an energy-averaged
transmission coefficient
\begin{equation}
\overline T_\pm^{\,{\rm rect \ bar}}\equiv \frac{1}{1+ \frac{1}{2}\,f_\pm(\overline E,U)}\cdot
\label{AvRectBarr}
\end{equation}
(the non-oscillating curve in Fig. \ref{fig: Fig5} and \ref{fig: Fig6}). 
As long as $\overline T_\pm^{\rm rect \ bar}$ is taken as an indicative prediction of the theory, the
seeming paradox shows up in the following way:
\begin{equation}
\overline T_\pm^{\, {\rm rect \ bar}}
\quad \stackrel{U\gg \overline E}\longrightarrow\quad 
\begin{cases}
0 \\ 
 \displaystyle{\frac{\overline E^{\,2}-m^2}{\overline E^{\,2}-m^2/2} }
\ \stackrel{\overline p^{\,2}\ll m^2}\longrightarrow\ 2\,\frac{\overline p^{\,2}}{m^2} 
\end{cases}
\label{highU} 
\end{equation} 
indicating an `unnaturally' high transmittivity of Dirac particles in presence of a high
step. The `unnatural' is referred to the comparison with the exponentially small Schr\"odinger
prediction. Even without going to the first limit taken in (\ref{highU}), for
example, for
$U=2.5$ MeV and average kinetic energy
$\overline E-m = 0.1$ MeV, the 31.3\% of the incident flux would be transmitted, the fraction going up
to  64.7\% for  
$\overline E-m = 0.5$ MeV.

The present paper addresses the question of whether Sauter's cure for the barrier is as good
as it is for the step. In addition to giving the analytical result for $T^{\rm Sauter \
barr}_{\rm D}$, our main result  consists in showing that, for the Sauter barrier, 
in the part of the Klein zone where $E-m\simeq m$,
\begin{equation}
\overline T^{\,\rm Sauter \ barr}_{\rm D}\propto \,(T^{\rm Sauter \ step}_{\rm
D})^2
\propto e^{-2\pi\, m^2\, \ell/U}
\label{TSauBar} 
\end{equation}
provided the slope $\ell$ is large enough so as to satisfy (\ref{Erestr}). 

As a matter of fact, for devices involving
either macro or mesoscopic dimensions, the result (\ref{TSauBar}) is a doom: for
practical purposes transmittivity is zero, in accordance with the naive
expectation.
Should one ever be able to set up a nanostructure with $\ell$ as small as
$10^{\,2
\div 3}
\lambda_{\rm Compton}$ and
$U$ not much higher than the threshold $2m$, it is seen that the constraint
(\ref{Erestr}) would be - even in this case - largely satisfied. 
Cases that could possibly be left uncovered 
by the present discussion
are those of solid state physics where the `effective mass' of the electron is smaller
than the in-vacuum-value adopted here. In particular the case of graphene, where a
vanishing effective mass is advocated by the authors of Ref. \cite{graphene}, looks to
date as the most promising ground where to observe the Klein paradox at work and, in the
most optimistic case, to set up a ``graphene device electronics" \cite{grapheneBis}.

The paper is organised as follows. 

In Section \ref{trmatr} the transfer matrix $M^{\rm step}$ for a generic
potential step is defined and the corresponding transmission coefficient discussed. In Section III the
tranfer matrix and transmission coefficient for the corresponding space-inversion invariant barrier are
expressed in terms of the matrix elements of
$M^{\rm step}$. Section IV (relying on the analytical results extracted from Sauter's original paper and
collected in the Appendix) illustrates the case of Sauter's trapezoidal barrier and justifies why the
averanging procedure described for (\ref{AvRectBarr}) applies and how is it that (\ref{TSauBar}) finally
comes about.


\section{Steps}    
\label{trmatr}

The present section contains material that is well known: as it mainly
serves to introduce our notation, some of the statements to be found below will be made
without proof. For the sake of conciseness we give a unified treatment  of the
Klein-Gordon equation in the first-order formalism and 
of  the Dirac equation, in one space dimension. In the latter case we make use of two-component
spinors since the spin - conserved by minimal coupling - is irrelevant. We will make use of
the Pauli matrices: 
\begin{equation*}
\sigma_1=\begin{pmatrix}0 & 1 \\ 1 & 0\end{pmatrix},\, 
\sigma_2=\begin{pmatrix}0 & -i \\ i & 0\end{pmatrix},\, 
\sigma_3=\begin{pmatrix}1 & 0 \\ 0 & -1\end{pmatrix} 
\end{equation*}
as well as of the $2\times 2$ identity matrix $I$.

We are interested in the stationary states
\begin{equation}
\Psi_E(x,t) = e^{-i\,E\,t}\, \Psi_E(x)\,,
\quad\Psi_E(x) = \begin{pmatrix}\psi_E(x)\\ \chi_E(x)\end{pmatrix}
\label{StSt} 
\end{equation}
of  the Klein-Gordon one-dimensional wave equation
\begin{equation}
\begin{cases}
\psi_E^\prime=\chi_E
\\
\chi_E^\prime = 
- \Big(\!\big(E-U(x)\big)^2-m^2\Big)\,\psi_E(x)
\end{cases}
\label{KGEq} 
\end{equation}
as well as of the Dirac equation. In the Pauli representation the
one-dimensional Dirac Hamiltonian is
\begin{equation}
H_{\rm D}^{\rm Pauli} =  -i\,\sigma_1\,\frac{d}{dx}+\sigma_3\,m+U(x)
\label{HD} 
\end{equation}
i.e., spelling out the components,  
\begin{equation}
\begin{cases}
\psi_E^\prime = i\,\big(E+m-U(x)\big)\,\chi_E
\\
\chi_E^\prime = i \, \big(E-m-U(x)\big)\,\psi_E\ .
\end{cases}
\label{DEq} 
\end{equation}
In both (\ref{KGEq}) and (\ref{DEq}) the energy  $E$ is understood to be in the Klein
zone (\ref{KZ}) and the step potential
$U(x)$ is given by
\begin{equation}
U(x) = \begin{cases}
0 & x<0 
\\
{\cal U}(x) & 0<x<\ell 
\\
U & x>\ell\ .
\end{cases}
\label{rectPot} 
\end{equation}
The solutions of the above equations may all be expressed in the form
\begin{equation}
\Psi_E = \begin{cases}
\alpha\,u(p)\,e^{i\,p\,x}+\beta\,u(-p)\,e^{-i\,p\,x} & x<0
\\
\Xi(x) & 0<x<\ell
\\
\gamma\,u(q)\,e^{i\,q\,x}+\delta\,u(-q)\,e^{-i\,q\,x} x>\ell 
\end{cases}
\label{stepSols} 
\end{equation}
where $\Xi(x)$, representing the solution in the intermediate region,
 evidently depends on the explicit form of ${\cal U}(x)$. The momenta $p$, $q$ are given by
(\ref{p}), (\ref{q}), and the two component wave functions are respectively 
\begin{align}
u_{\rm KG}(\pm p) = \begin{pmatrix} 1\\ \\ \pm i \,p\end{pmatrix} &\,, 
\quad
u_{\rm D}(\pm p) = \begin{pmatrix} 1 \\ \\ \displaystyle\frac{\pm p}{E+m}\end{pmatrix}
\\ 
u_{\rm KG}(\pm q) = \begin{pmatrix} 1\\ \\ \pm i \,q\end{pmatrix} &\,,
\quad
u_{\rm D}(\pm q) = \begin{pmatrix} 1 \\ \\ \displaystyle{\frac{\pm q}{E-U+m}}\end{pmatrix}\!\cdot
\label{spinors2} 
\end{align} 
Imposing the continuity of $\Psi_E$ at the points $x=0,\,x=\ell$ entails two relations of
linear dependence among the four constants $\alpha,\,\beta,\,\gamma,\,\delta$.
The transfer matrix $M^{\rm step}$, expressing such a dependence,
is defined by
\begin{equation}
\begin{pmatrix}\gamma \\ \delta \end{pmatrix} 
= M^{\rm step} \, 
\begin{pmatrix}\alpha \\ \beta \end{pmatrix}
= \begin{pmatrix} a & b \\ 
c & d\end{pmatrix}
\begin{pmatrix}\alpha \\ \beta\end{pmatrix}\cdot
\label{SauterM} 
\end{equation}

Use of the invariance of 
(\ref{KGEq}) and (\ref{DEq}) under charge conjugations
\begin{equation}
{\cal C}: \quad \Psi(x) \to C\, \Psi^*(x)\,,
\qquad
C=\begin{cases}
I & \quad \text{KG} 
\\ 
\sigma_3 & \quad \text{D}
\end{cases}
\label{CC} 
\end{equation}
simplifies the form of $ M^{\rm step}$ down to
\begin{equation}
 M^{\rm step}= 
\begin{pmatrix}
a & b \\ b^* & a^*
\end{pmatrix}\cdot
\label{Mstep}
\end{equation}
In addition, conservation of the currents (whose form is dictated by Noether's theorem), i.e. the
constance of
\begin{equation}
J_x = 
\begin{cases} 
\Psi^\dagger\,\sigma_2\,\Psi &\qquad \text{KG} \\
\Psi^\dagger\,\sigma_1\,\Psi &\qquad \text{D}
\end{cases}
\label{curr} 
\end{equation}
with respect to $x$, implies
\begin{equation}
|\alpha|^2-|\beta|^2 = {1\over {\cal D}_\pm}\, \frac{q}{p}\,(|\gamma|^2-|\delta|^2)
\label{fluxCons} 
\end{equation}
where
\begin{equation}
{\cal D}_\pm = 
\begin{cases}
1 & \qquad += \text{KG} 
\\ \\
\displaystyle{\frac{E-U+m}{E+m}} & \qquad -= \text{D}\ .
\end{cases}
\label{Ratio} 
\end{equation}
Finally, expressing $\gamma,\,\delta$ in (\ref{fluxCons}) as given by
(\ref{SauterM}), one obtains
\begin{equation}
\det M^{\rm step}_\pm = |a|^2-|b|^2=\frac{p}{q} \ {\cal D}_\pm\ .
\label{dets}
\end{equation}
In the Klein-Gordon case there is no doubt that the scattering state with the $\alpha$ source term
to the left is identified by setting $\delta=0$ in
(\ref{stepSols}). Indeed, according to (\ref{fluxCons}), only an 
outgoing (from left to right) flux is left in the region $x>\ell$.
The transmission coefficient is therefore, in general
\begin{equation}
T^{\rm step}_{\rm KG} \equiv \frac{q}{p}\ \left\vert\frac{\gamma}{\alpha}\right\vert^2
=\frac{p}{q}\ {1\over |a|^2}\ \cdot
\label{TKGstep} 
\end{equation}
Apparently Pauli was the first \cite{KL} to note the impact that
${\cal D}_-<0$ has, in the Klein zone, on the identification of the scattering states.
As the contributions to the flux of Dirac particles in the right end side of (\ref{fluxCons}) are
interchanged with respect to the Klein-Gordon case,
the transmission coefficient should be obtained by setting $\gamma=0$ instead of $\delta$ in 
(\ref{stepSols}). This choice gives, for the generic form ${\cal U}(x)$ of the step,
\begin{equation}
T^{\rm step}_{\rm D} \equiv \frac{q}{p}\ \frac{E+m}{U-E-m } \
\left\vert\frac{\delta}{\alpha}\right\vert^2 =-\,\frac{p}{q}\    {{\cal D}_- \over |b|^2}
\label{TDstep} 
\end{equation}
to be contrasted with (\ref{TKGstep}). 

In the case of the rectangular step, explicit calculation yields
\begin{equation}
a_\pm^{\rm rect \ step}= \frac{1}{2} \Big(1+\frac{p}{q} \ {\cal D}_\pm\Big)\, ,
\qquad  
b_\pm^{\rm rect \ step}= \frac{1}{2} \Big(1-\frac{p}{q} \ {\cal D}_\pm\Big)
\label{abD} 
\end{equation}
whence
\begin{equation}
T^{\rm rect \ step}_{\pm}=  \frac{\displaystyle {\pm \ 4\,\frac{p}{q}\,{\cal D}_\pm }}{\displaystyle {\left(1
\pm \frac{p}{q}{\cal D}_\pm\right)^2}} \ \cdot
\label{Trecsteps} 
\end{equation}

In the Dirac case it is evident that our
identification of the scattering states 
is in disagreement, e.g., with the choice of Ref.
\cite{BD}, where the interchange of the
$\gamma$ and $\delta$ terms is not effected.
Our result is indeed obtained from theirs by effecting the substitution $r\to -r$ (our $p/q\ {\cal D}_-$
equals Bjorken-Drell's  $1/r$), which turns their transmission coefficient - negative in the Klein zone -
into our positive expression.

Considering now that, for fixed $E$,
\begin{equation}
\lim_{U\to\infty}\ -\frac{p}{q}\ {\cal D}_-= \sqrt{\frac{E-m}{E+m}} 
\end{equation}
it follows that
\begin{equation}
\lim_{U\to\infty}\ T^{\rm rect \ step}_{\rm D}=\frac{2p}{E+p} 
\end{equation}
i.e. a {\it finite} value, to be contrasted with the Klein-Gordon case where
\begin{equation}
\lim_{U\to\infty}\ \frac{p}{q}= 0 
\end{equation}
entails
\begin{equation}
\lim_{U\to\infty}\ T^{\rm rect \ step}_{\rm KG}=0\,. 
\end{equation}

\begin{figure}
              \includegraphics{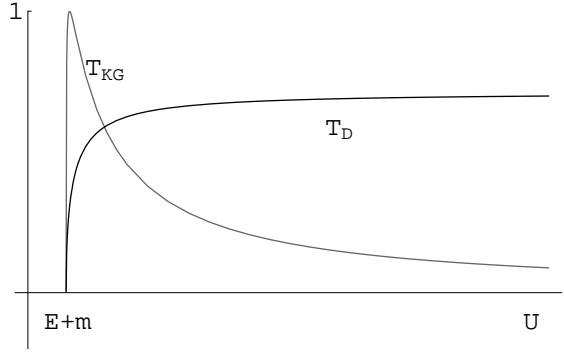}
              \vspace{.1 cm}
\caption{$T^{\rm rect \, step}_{\rm D}$  and $T^{\rm rect \, step}_{\rm KG}$  for $E=1.2\,m$
as functions of $U$ in the Klein zone} \label{fig: Fig7}
\end{figure}

The situation is summarized in Fig. \ref{fig: Fig7}.
The comparison of the two cases is self evident. Whereas, in the first case, the barrier is transparent
for essentially all the values of the potential, in the second one, apart from a region just above
the threshold of the Klein zone, the barrier becomes again impenetrable. This shows how wrong is the naive
idea that only the oscillatory behaviour of the eigenfunction - due to the relativistic energy-momentum
relation (\ref{q}) - does all the job in the Klein zone. Indeed the equation for the large
component of the Dirac equation (obtained by taking the derivative of the first equation in (\ref{DEq})
and there using the second)
is not equivalent to the Klein-Gordon equation (\ref{KGEq}). The derivative, acting on $U(x)$, yields extra
terms $e\,{\cal E} = - U'(x)$, thus supporting the expectation that a sharp edge is fundamental in
making the difference between the previous results. This is indeed the content of Sauter's work \cite{SA}.

At this stage we can no longer help interpreting the scattering states. Indeed, in the scattering state
identified by choosing
$\gamma=0$ in (\ref{stepSols}), the positive (i.e. toward the
right) {\it current} in the region $x>0$ cannot come from electrons with negative kinetic energy, see
(\ref{electrons}). It is rather due to positrons propagating backward in time, see viceversa
(\ref{positrons}). Hence - in the
language of second quantization - a physical positron above the potential
$-U$ comes from the right, i.e the $\delta$ term acts as a source of positrons available at $x=+\infty$.  In
this sense we are, therefore, in the presence of a $e^+$-$e^-$ annihilation process, and the reflection
coefficient $R=1-T$ {\it necessarily} is smaller than 1.  Had we had alternatively chosen $\delta=0$ in
(\ref{stepSols}), we would have had, also from the right, an incoming current with an ensuing reflection
coefficient bigger than 1. This would correspond to a flux of positrons travelling to the right, i.e. a
process of pair creation. 
And this is the result much quoted as the `Klein paradox', a statement that simply is not present in the
original Klein's paper. Let us however again stress that even for $R<1$ there is a paradox due to a finite 
fermion transmission through a high repulsive potential. Analogous considerations apply in the Klein-Gordon
case with the previous provisions (Eq.s (\ref{TKGstep}) and (\ref{abD}) and Fig. \ref{fig: Fig7}) due
to the reversed role of the
$\gamma$ and $\delta$ terms.

\section{Barriers}
\label{bars} 

The space-inversion invariant potential barrier obtained by using twice the step (\ref{rectPot}) is
\begin{equation}
W(x) = U\big(x+\ell+L \big) + U\big(-(x+\ell+L)\big)
\label{BarrPot} 
\end{equation}
and the corresponding stationary states have the form
\widetext
\begin{equation}
\Psi_E(x)=\begin{cases}
\alpha\,u(p)\,e^{i\,p\,x}+\beta\,u(-p)\,e^{-i\,p\,x} & x<-(\ell+L)
\\
\Xi_-(x) & -(\ell+L) < x< -L  
\\
\gamma\,u(q)\,e^{i\,q\,x}+\delta\,u(-q)\,e^{-i\,q\,x} &-L<x<L
\\
\Xi_+(x) & L < x< \ell+L  
\\
\alpha^\prime\,u(p)\,e^{i\,p\,x}+\beta^\prime\,u(-p)\,e^{-i\,p\,x} &x>\ell+L
\end{cases}
\label{barSols} 
\end{equation}
  
\noindent
(the coefficients $\alpha,\,\beta,\,\gamma,\,\delta$ appearing in the latter equation
should not be confused with those appearing in (\ref{stepSols}) to which they
are however related through the translation $x\to x+\ell+L$). Again, the continuity
relations at
$x=-(\ell+L),\,-L$ and 
$x=L,\,\ell+L$ respectively entail
\begin{equation}
\begin{pmatrix}
\gamma\\ \delta
\end{pmatrix} 
= M^{\rm left} \, 
\begin{pmatrix}
\alpha \\ \beta
\end{pmatrix}\,,
\qquad
\begin{pmatrix}
\alpha^\prime \\ \beta^\prime
\end{pmatrix} 
= M^{\rm right} \, 
\begin{pmatrix}
\gamma \\ \delta
\end{pmatrix}
\label{Mlr} 
\end{equation} 
whence the transfer matrix relative to the barrier
\begin{equation}
\begin{pmatrix}
\alpha^\prime\\ \beta^\prime
\end{pmatrix}
\equiv M^{\rm bar} 
\begin{pmatrix}
\alpha\\ \beta
\end{pmatrix}
=
\begin{pmatrix}
A & B \\  B^* & A^*
\end{pmatrix}
\begin{pmatrix}
\alpha\\ \beta
\end{pmatrix}
\label{Mbar} 
\end{equation}
is obtained as
\begin{equation}
M^{\rm bar}= M^{\rm right}M^{\rm left}\ .
\label{Mlefrig} 
\end{equation}
The connection of $M^{\rm right}$ and  $M^{\rm left} $ with $M^{\rm step}$
is established by taking translations and reflection into account.
The effect of the translation to the left $x\to x+\ell+L$ is expressed
with the aid of the matrix
\begin{equation}
T(k,a) = 
\begin{pmatrix}
e^{i\,k\,a} & 0 
\\
0 & e^{-i\,k\,a}
\end{pmatrix}
\cdot
\label{T} 
\end{equation} 
One gets
\begin{equation}
M^{\rm left} = 
T(q,\ell+L)\, 
M^{\rm step}\,
T(-p,\ell+L)
\label{Mleft} 
\end{equation}
whereas $M^{\rm right}$ is obtained either by reflecting $x\to-x$ in $M^{\rm left}$,
or by first reflecting $M^{\rm step}$ and then translating to the
right
$x\to x-(\ell+L)$:
$$
M^{\rm right} = \sigma_1\,(M^{\rm left})^{-1}\,\sigma_1=
$$
\begin{equation}
T\big(p,-(\ell+L)\big)\,
\sigma_1(M^{\rm step})^{-1}\sigma_1\,
T\big(-q,-(\ell+L)\big)\, .
\label{Mright} 
\end{equation}
Exploiting (\ref{Mlefrig})-(\ref{Mright}) one finally obtains for
the coefficients appearing in (\ref{Mbar})
\begin{equation}
A = {a^2 \,e^{2i(q-p)(\ell+L)}-b^{*\,2}\,e^{-2i(q+p)(\ell+L)} \over |a|^2-|b|^2 }
\label{A} 
\end{equation}
\begin{equation}
B = {a\,b \,e^{2iq\,(\ell+L)}-a^*\,b^*\,e^{-2iq\,(\ell+L)}\over |a|^2-|b|^2 }\cdot
\label{B} 
\end{equation}
In the case of the barrier we do not have the ambiguity - connected with the identification of the
scattering states - we have discussed for the step in (\ref{TKGstep}) and (\ref{TDstep}).
The transmission coefficient of the barrier is 
$1/|A|^2$ both in the Klein-Gordon and the Dirac case. When it is expressed in terms of the two matrix
elements entering the transfer matrix of the step
\begin{equation}
a = |a| \, e^{i\,\xi}\,,\quad b = |b|\, e^{i\,\eta}\,,
\label{ab} 
\end{equation}
takes the form
\begin{equation}
T^{\rm bar}\! =\! \frac
{1}
{1+ \displaystyle{
\frac{4|a|^2|b|^2}
{(|a|^2\!-\!|b|^2)^2}
\sin^2\!\big(2q\,(\ell+L)+\xi +\eta\big)}}
\cdot
\label{Tbar} 
\end{equation}
This can be further simplified on account of (\ref{dets})-(\ref{TDstep}):
\begin{equation}
T^{\rm bar}\! =\! \frac{1}{1+ 4\,\displaystyle{\frac{1-T^{\rm step}}{(T^{\rm step})^2}\,
\sin^2\big(2q\,(\ell+L)+\xi +\eta\big)}}\cdot
\label{TbarD} 
\end{equation}
The above formula is consistent with the results cited in the introduction, i.e. when
(\ref{TKGstep})-(\ref{Trecsteps}) - appropriate for the rectangular barriers - are used,
(\ref{rectBarr}), (\ref{KGD}) are reobtained.

The dependence of $T^{\rm bar}$ on the energy is through the phases $\xi$, $\eta$ of the transfer matrix
elements (\ref{ab}), as well as through
$T^{\rm step}$ itself and $q$. The dependence on the  barrier width $2L$ is, instead, completely spelled out
in the argument of the
$\sin$ function. This means that for given energy, no matter
how opaque the step ${\cal U}(x)$ may be, the width parameter $L$ can be always adjusted in such a way that
the corresponding even barrier is totally transparent (this is analogue to the situation of
dielectric stratified media in optics: see e.g. \cite{BW}, where however only the case of sharp edge
$\ell=0$, i.e. $\xi =
\eta=0$, is considered). Going back to particles, as discussed in the introduction, there may be the
difficulty that,
 for given average energy $\overline E$ and energy uncertainty $\Delta E$ of the incident
particles, any `realistic' value of $L$ may be such that too many maxima of $T^{\rm bar}$, due to the
oscillations of $\sin^2(\dots)$, come into the range ($\overline E-{1\over 2}
\Delta E,\overline E+{1\over 2}
\Delta E$). The averaging procedure $\sin^2(\dots)\to {1\over 2}$ in (\ref{TbarD}) is then necessary
and (\ref{TSauBar}) follows immediately when $T^{\rm step}\ll 1$.

\section{Sauter trapezoidal Barrier for the Dirac case}

The analytic expression for $T^{\rm Sauter \ bar}_{\rm D}$ can be deduced
from the long formulae we have extracted from Sauter's paper \cite{SA} and reported in the Appendix.
The graphs below are more eloquent than the expression.
In assigning numerical values to the parameters, we have in mind the case
of electrons in vacuum. Therefore the unit of energy is $m=.51$ MeV and the unit of lenght is
$\lambda_{\rm Compton}=.024$ \AA.

We repeat here the warning already made in the Introduction: there may be cases where the units
change by orders of magnitude and the following discussion may not apply.

In consequence of the scales we have chosen, the values we give to $U$ are not much higher than the
threshold, typically
$U=3\,m$. As for
$\ell$ and
$L$, certainly they both have to take a value of the order of at least $10^2\,\lambda_{\rm Compton}$. It is
however instructive to vary one parameter at a time.

\begin{figure}
              \includegraphics{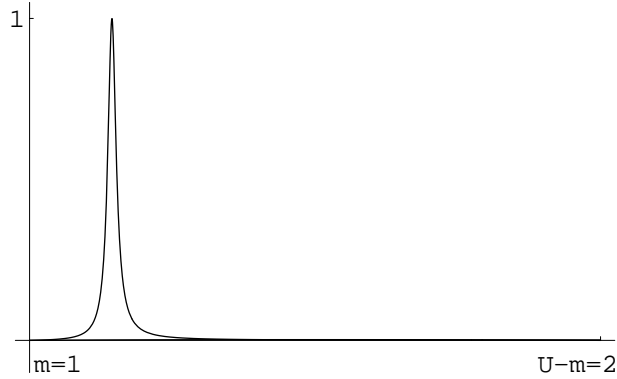}
              \vspace{.1 cm}
\caption{$T^{\rm Saut \, bar}_{\rm D}$    for $U=3\,m,\ L=0, \ \ell=3\,\lambda_{\rm Compton}$
as function of $E$ in the Klein zone} \label{fig: Fig8}
\end{figure}

Fig. \ref{fig: Fig8}  plots $T^{\rm Saut \, bar}_{\rm D}$ against the energy $E$ in the Klein zone for
$U=3m$, $\ell = 3\lambda_{\rm Compton}$ and $L=0$: it is a triangular barrier that shows the effect of
$\ell$ alone. It indicates that a flux of monoenergetic electrons
($\Delta E
\ll$ width of the peak) is entirely transmitted. This may come a little bit as a surprise because,
referring to Fig. \ref{fig: Fig2} and in a
pictorial  space-time description, the incoming electron encounters:

(i) 
a free propagation region (from $x=-\infty$ to $x=0$); 

(ii)
a region from $0$ to $x_1$ that is still classically accessible, but where the propagation
is non-free:    the two solutions behave as
(think e.g. of the semiclassical approximation) exponentials of opposite imaginary arguments; 

(iii)
the classically inaccessible region
$(x_1,x_2)$ with non-propagating solutions: the two solutions of the Dirac equation behave as
exponentials of opposite real arguments (a region that we will improperly call the `attenuation' region);

(iv) eventually  the classically inaccessible region from $x_2$ to $\ell$ where propagation, albeit 
non-free, still occurrs;

\noindent
then, from $\ell$ to $+\infty$, the same steps just described, but in reverse order. 

If one takes the Schr\"odinger case as
a guidance, one might expect attenuation of the wave function both in the left and in the right attenuation
region.  This  happens for almost all the values of $E$ in the Klein-zone, with the exception of the peak
displayed. Indeed, for average energy $\overline E$ at the center of the peak, the single monocromatic
components, that make up the incident packet, undergo two transmissions at the edges and possible multiple
reflections in between, but they succeed in keeping somehow memory of the relative phases and the packet is
(almost entirely) reconstructed beyond the barrier. For the values we have indicated the `device' could serve
as a monocromator for incident wave packets with energy uncertainty greater than the width of the displayed
peak. From a practical point of view this is illusory, due to the exceedingly small value of $\ell=
3\lambda_{\rm Compton}= 0.072$ \AA. 

\begin{figure}
              \includegraphics{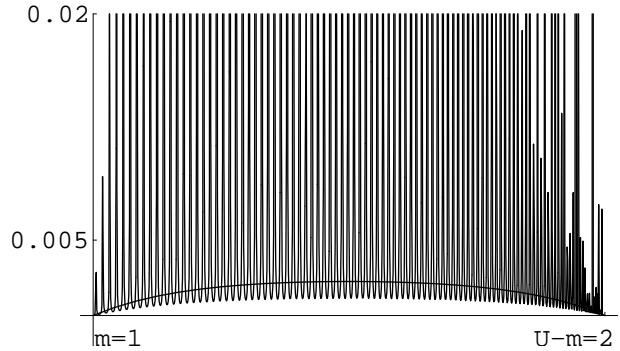}
              \vspace{.1 cm}
\caption{$T^{\rm Saut \, bar}_{\rm D}$    for $U=3\,m,\ L=100\,\lambda_{\rm Compton}, 
\ell=$ $3\,\lambda_{\rm Compton}$ as function of $E$ in the Klein zone} \label{fig: Fig9}
\end{figure}

Fig. \ref{fig: Fig9} shows the impact of increasing $L$ from 0 to  $100\, \lambda_{\rm Compton}=2.4\,$ 
\AA  in the preceding case. A region of strict free propagation is interposed, the peak of Fig.
\ref{fig: Fig8} is replicated several times (the argument of the $\sin^2(\dots)$ in (\ref{TbarD})
oscillates much faster) and the peaks displayed all reach the value 1.
However we have chosen to reduce the scale of the ordinates
to make the energy averaged
$\overline T^{\rm Saut \, bar}_{\rm D}$ visible. The oscillations of the non-averaged $T$ are so fast (and
still the value given to $L$ may be largely considered too small) that only $\overline T$ is susceptible of
comparison with the data of whatever `Gedanken' experiment. Nonetheless a value of $\overline T$ of the
order of some part per thousand could still be considered an interesting signal.

\begin{figure}
              \includegraphics{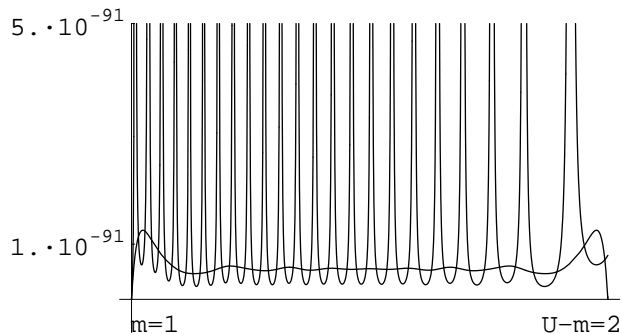}
              \vspace{.1 cm}
\caption{$T^{\rm Saut \, bar}_{\rm D}$    for $U=3\,m,\ L=0, \ \ell=100\,\lambda_{\rm Compton}$
as function of $E$ in the Klein zone} \label{fig: Fig10}
\end{figure}

Once the role of $L$ is understood,
in Fig. \ref{fig: Fig10} we go back to the case $L=0$ but raise the value of $\ell$ from $3\,\lambda_{\rm
Compton}$ to $100\,\lambda_{\rm
Compton}$. The main difference is in the scale of the ordinates. Again the peaks all reach the value 1,
but the scale of $\overline T$ has dropped to $10^{-90\div 91}$. As for the part of the Klein zone that
also satisfies $E-m\simeq m$, this is in perfect agreement with (\ref{TSauBar})).

The analytical result, displayed in the figure, shows that the
estimate extends to the whole Klein zone.  

The number of peaks in the Klein zone increases with increasing
$\ell$, but the width of each of them shrinks much faster: this is why the
value of
$\overline T$ has dropped of so many orders of magnitude.

Turning $L$ on will not appreciably change the scale of the ordinates, but the energy averaging becomes
unavoidable. This washes out any phase information among the monoenergetic components the incident packet is
made of. As a consequence the packet, that has undergone an attenuation in the attenuation region at the
first edge, may only undergo a second attenuation at the second edge. The result  is that the transmission
coefficient (\ref{TbarD}) is shattered down to the asymptotic value given in (\ref{TSauBar}).

Finally the trapezoidal form of the barrier, we have chosen in order to be able to exhibit
analytical results, should not be crucial to the above conclusion. The same qualitative result should obtain
with all the barriers where the dimension $\ell$ of the regions that support a relevant electric field ($e\,
{\cal E} > 2m^2)$ is large enough so as to fulfill (\ref{Erestr}). Ref. \cite{Z2}, where again Sauter
provides analytical results for the potential
\[U_1(x)= {U\over 2}\left(1+ \tanh {x\over \ell}\right)\]
as well as the relevant asymptotics, substantiates the above expectation.

The foregoing argument should also work for the `delocalization' of particles in supercritical potentials
\cite{GZ}, \cite{GZG}. This will be possibly considered elsewhere.

\acknowledgments

The authors are thankful for the enlightening comments by
P. Menotti. They have also benefitted by 
a stimulating discussion with S. De Leo and P. Rotelli. 

\appendix* 


\section{Linearly raising potential and Sauter solution for the Dirac Case} 
\label{Sauter}

Referring to the Dirac equation with the step potential of Fig. \ref{fig: Fig2}
\begin{equation}
{\cal U}(x) = U\,{x\over \ell}\ \,, \qquad 0\le x \le \ell
\end{equation}
the solution $\Xi(x)$ in the intermediate region $0<x<\ell$ to be fed in (\ref{stepSols}) can be
worked out from Sauter's article \cite{SA}. His representation for the Dirac Hamiltonian is different from
Pauli's (\ref{HD}): he uses
\begin{equation}
 H_{\rm D}^{\rm Sauter}=  i\,\sigma_3 \frac{d}{dx}+U(x)-\sigma_2\,m= V\,H_{\rm D}^{\rm Pauli}\,
V^\dagger
\end{equation}
with (up to an irrelevant phase factor)
\begin{equation}
V = \frac{1}{2}\big(I-i\,(\sigma_1+\sigma_2-\sigma_3)\big)\ .
\label{V}
\end{equation}
He then introduces the argument
\begin{equation}
\xi(x)=\sqrt{\frac{U}{\ell}}\Big(x- {E\over U}\, \ell \Big)
\end{equation}
and the degenerate hypergeometric function
\begin{equation}
\Phi(\alpha,\beta;z)= 1 + \frac{\alpha}{\beta}\, \frac{z}{1!}+\frac{\alpha(\alpha+1)}{\beta(\beta+1)}\,
\frac{z^2}{2!}+\dots
\end{equation}
in terms of which the two independent eigenfunctions belonging to energy $E$
\begin{equation}
\Sigma=
\begin{pmatrix}
f_E(x)\\  g_E(x)
\end{pmatrix} \,,\quad
\widetilde\Sigma = {\cal C}\Sigma= \sigma_1\Sigma^*=
\begin{pmatrix}
g_E^*(x)\\  f_E^*(x)
\end{pmatrix}
\end{equation} 
in the intermediate region $0<x<\ell$ are  given by
\begin{align}
f_E(x) & =  e^{i\, \xi^2/2}\,\Phi\left(\frac{i}{4}\,\frac{m^2\,\ell}{U},\frac{1}{2},-i\,\xi^2\right)
\\
g_E(x) & =  -m\,\sqrt{\frac{\ell}{U}}\,e^{i\, \xi^2/2} \,
\Phi\left(\frac{i}{4}\,\frac{m^2\,\ell}{U}+1,\frac{3}{2},-i\,\xi^2\right) .
\end{align}
Thanks to (\ref{V}),  the $\Xi(x)$ to be fed in (\ref{stepSols})  is the linear combination
\begin{equation}
\mu
\begin{pmatrix}
f_E(x)+ig_E(x)
\\ 
-f_E(x)+ig_E(x)
\end{pmatrix} 
+ \nu\,
\begin{pmatrix}
g_E^*(x)+if_E^*(x)
\\ 
-g_E^*(x)+if_E(x)
\end{pmatrix}\,.
\end{equation}
The  transfer matrix $M^{\rm Saut \ step}_{\rm D}$ is obtained by imposing the continuity of
(\ref{stepSols}) at
$x=0$ and $x=\ell$ and then eliminating $\mu$ and $\nu$. Its matrix elements (\ref{Mstep}) turn out
to be 
\goodbreak
\begin{align}
a^{\rm Saut \ step}& = \frac{e^{-i\,q\,\ell}}{4q\,(E+m)}
\
 \Big(P_+Q_+R-P_-Q_-R^*+i\,\big(P_+Q_-S+P_-Q_+S^*\big)\Big)
\\ 
b^{\rm Saut \ step}&= \frac{e^{-i\,q\,\ell}}{4q\,(E+m)}
\
\Big(P_-Q_+R-P_+Q_-R^*+  i\,\big(P_-Q_-S+P_+Q_+S^*\big)\Big)
\end{align}
where
\begin{align}
R&=f_E^{\phantom{*}}(0)\,f_E^*(\ell) - g_E^*(0)\,g^{\phantom{*}}_E(\ell)
\\
S&= f_E^{\phantom{*}}(0)\,g_E^*(\ell)- g_E^{\phantom{*}}(0)\,f^*_E(\ell)
\\
P_\pm&=E+m\pm p
\\
Q_\pm&=E-U+m\pm q\ .
\end{align}
Substitution of the above formulae in (\ref{ab})-(\ref{TbarD}) provides the
analytic expression for $T^{\rm Saut \ step}_{\rm D}$ by which the  graphs
of Fig.s \ref{fig: Fig8}-\ref{fig: Fig10} above are obtained. Finally Sauter himself
works out the asymptotic form (\ref{TSau}) that $T^{\rm Saut \ step}_{\rm D}$ has when both $E-m\simeq m$
and (\ref{Erestr}) are satisfied:
\[
T^{\rm Saut \ step}_{\rm D} = {q\,(U-E+q)\over p\,(E+p)}\, e^{-\pi \, m^2\ell/U}\,\big(1+ {\cal
O}(U/m^2\ell)\big)
\]
having included a prefactor omitted in (\ref{TSauBar}).

\goodbreak


\goodbreak
\vfill\eject
\end{document}